# On the Powerball Method for Optimization

Ye Yuan[a], Mu Li[b], Jun Liu[c], Claire Tomlin[d]

[a]*School of Automation, Huazhong University of Science and Technology*
[b]*Department of Computer Science, Carnegie Mellon University*
[c]*Department of Applied Mathematics, University of Waterloo*
[d]*Department of Electrical Engineering and Computer Sciences, University of California, Berkeley*

**Abstract**

We propose a new method to accelerate the convergence of optimization algorithms. This method simply adds a power coefficient $\gamma \in [0, 1)$ to the gradient during optimization. We call this the Powerball method and analyze the convergence rate for the Powerball method for strongly convex functions. While theoretically the Powerball method is guaranteed to have a linear convergence rate in the same order of the gradient method, we show that empirically it significantly outperforms the gradient descent and Newton's method, especially during the initial iterations. We demonstrate that the Powerball method provides a 10-fold speedup of the convergence of both gradient descent and L-BFGS on multiple real datasets.

*Keywords:* Optimization algorithms, Convex programming, Dynamical systems, Lyapunov function, Convergence analysis

## 1. Introduction

We consider minimizing a differentiable function $f(x) : \mathbb{R}^n \to \mathbb{R}$ with iterative methods. Given a starting point $x(0) \in \mathbb{R}^n$, these methods compute

$$x(k+1) = x(k) - A_k^{-1} \nabla f(x(k)), \quad \text{for } k = 0, 1, \dots. \quad (1.1)$$

Previous work has focused mainly on the choice of $A_k$. One choice is using a scalar step size $A_k = \alpha_k^{-1}$ with $\alpha_k > 0$, yielding the gradient descent method due to Cauchy. Another widely adopted choice of $A_k$ is the Hessian matrix $\nabla^2 f(x(k))$, which is used by the notable Newton's method.

In this paper, we propose the Powerball method, which applies a nonlinear element-wise transformation to the gradient by

$$x(k+1) = x(k) - A_k^{-1} \sigma_\gamma(\nabla f(x(k))), \quad \text{for } k = 0, 1, \dots. \quad (1.2)$$

For any vector $x = (x_1, \dots, x_n)^T$, the Powerball function $\sigma_\gamma$ is applied to all elements of $x$, that is $\sigma_\gamma(x) = (\sigma_\gamma(x_1), \dots, \sigma_\gamma(x_n))^T$. For simplicity, we drop the subscript $\gamma$ and use $\sigma(x)$ to denote $\sigma_\gamma(x)$. The Powerball function $\sigma(\cdot) : \mathbb{R} \to \mathbb{R}$ has the form $\sigma(z) = \text{sign}(z)|z|^\gamma$ for $\gamma \in (0, 1)$, where $\text{sign}(z)$ returns the sign of $z$, or 0 if $z = 0$. We use a constant power coefficient $\gamma$ for all iterations. Similarly, we call the method with $A_k = \alpha_k^{-1}$ in eq. (1.2) the gradient Powerball method and the method with $A_k = \nabla^2 f(x(k))$ the Newton Powerball method. We will also propose other Powerball variants of standard methods throughout the paper, for example, the L-BFGS Powerball method.

This paper is organized as follows. In Section 2, we shall provide intuition behind the Powerball method by viewing optimization algorithms as discretizations of ordinary differential equations (ODE). Furthermore, we analyze the convergence rate for the proposed Powerball method for strongly convex functions in Section 3 and discuss important variants of Powerball method in Section 4. Moreover, we demonstrate the fast convergence of Powerball algorithms on a classification problem with benchmark datasets in Section 5. Finally, we conclude this paper with a general discussion on applying insights from control and dynamical systems to optimization algorithms.

## 2. Intuition from Ordinary Differential Equations

Consider the algorithms presented in eq. (1.1) and eq. (1.2). If the index, or iteration number, of these algorithms is viewed as a discrete-time index, then these algorithms can be viewed as discrete-time dynamical systems. By taking this view, the convergence of an optimization method to a minimizer can be equivalently seen as the convergence of a dynamical system to an equilibrium (Bhaya and Kaszkurewicz, 2006).

The intuition of the gradient Powerball algorithm lies in the Euler discretization of the following ODE:

$$\dot{x} = -\sigma(\nabla f(x)). \quad (2.1)$$

**Definition 1.** *A function $f$ is strongly convex with coefficient $m > 0$, if it satisfies $f(y) \geq f(x) + \nabla f(x) \cdot (y-x) + \frac{m}{2}\|y-x\|^2$ for all $x, y \in \mathbb{R}^n$.*

We prove a convergence result for the above ODE when $\gamma \in (0, 1)$ under the assumption that $f$ is strongly convex. The proof is given in Appendix A.

**Proposition 1.** *For any strongly convex function $f$ with coefficient $m$, the solutions of the ordinary differential equation (2.1) for $\gamma \in (0, 1)$ converge to its equilibrium in finite time $T = \frac{((\gamma+1)V(0))^{\frac{1-\gamma}{1+\gamma}}}{m(1-\gamma)}$, in which $V(t) \triangleq \frac{1}{\gamma+1} \sum_{i=1}^{n} \left|\frac{\partial f(x(t))}{\partial x_i}\right|^{\gamma+1}$.*

Similarly, the intuition of the Newton Powerball method lies in the Euler discretization of the following ODE:

$$\dot{x} = -\left(\nabla^2 f(x)\right)^{-1} \sigma(\nabla f(x)).$$

**Proposition 2.** *For any twice differentiable function $f$, the proposed continuous Newton Powerball method converges to an equilibrium point in finite time $T = \frac{((\gamma+1)V(0))^{\frac{1-\gamma}{1+\gamma}}}{1-\gamma}$, in which $V(t) \triangleq \frac{1}{\gamma+1} \sum_{i=1}^{n} \left|\frac{\partial f(x(t))}{\partial x_i}\right|^{\gamma+1}$.*

*Proof.* See Appendix B. □

Through analyzing the continuous versions of optimization algorithms and viewing convergence of continuous optimization algorithms as stability of dynamical systems, we can apply Lyapunov theory from control theory to gain insight about the underlying optimization algorithms. What remains is to derive an analogous proof for discrete-time dynamical systems, or equivalently for optimization algorithms. As pointed out by Su, Boyd and Candes (Su at al., 2015), the translation of ODE theory to optimization algorithms involves parameter tuning (for example, step-size) and tedious calculations. We shall derive, in the following section, the convergence rate for Powerball methods for strongly convex functions so that we can compare it with rates for standard methods.

## 3. Convergence Analysis

Given the intuition in the previous section, we propose the gradient Powerball method whose the iterative scheme writes

$$x(k+1) = x(k) - \alpha_k \sigma(\nabla f(x(k))), \quad \text{for } k = 0, 1, \ldots, \quad (3.1)$$

where $\alpha_k$ is the step size to be chosen. We shall show the convergence for Powerball methods in eq. (1.2) for strongly convex functions. The proof is given in Appendix C.

**Theorem 1.** *For any strongly convex function $f$ (with coefficient $m$) with $L$-Lipschitz gradient, the proposed gradient Powerball method converges at least linearly to the global minimizer at a rate $(1 - O(\frac{m}{L}))^k$.*

**Remark 1.** *While the proved linear convergence rate is in the same order as the standard gradient descent method, the Powerball method seems to outperform the standard gradient descent method, as demonstrated in the examples in Section 5. We note that, in its essence, the Powerball method can be regarded as the steepest gradient descent with respect to the p-norm, where $p = 1 + \frac{1}{\gamma}$ (cf. Section 9.4 of Boyd and Vandenberghe (2004)). Here, $\gamma$ serves as an additional parameter for tuning the performance of the proposed scheme.*

**Remark 2.** *The step size $\alpha_k$ can be chosen in a specific way as in the proof. We can also apply standard backtrack line search to achieve potentially better empirical performance in practice.*

**Remark 3.** *It remains an interesting open theoretical question to derive a convergence rate that a) exhibits the finite-time convergence property we observed for its continuous-time version; b) explicitly depends on the parameter $\gamma$ and; c) in particular, explains the empirical speedup during the initial iterations, as observed in Section 5.*

## 4. Variants of the Powerball Method

In this section, we consider the following two variants of the proposed Powerball method.

### 4.1. One-bit gradient descent method

First, it is natural to consider the special case $\gamma = 0$, which has a very low communication cost for optimizing strongly convex functions: it reduces the communication bandwidth requirement for the data exchanges (Seide et al., 2014) since only the sign for every element of the gradient computation is needed. The one-bit gradient descent method has the following form (simply let $\gamma = 0$)

$$x(k+1) = x(k) - \alpha_k \text{sign}(\nabla f(x(k))) \quad \text{for} \quad k = 0, 1, \ldots. \quad (4.1)$$

### 4.2. L-BFGS Powerball method

The L-BFGS method is a quasi-Newton method (Yuan, 2011) which achieves a similar convergence rate as Newton's method near the optimal solution. L-BFGS is widely used in practice. We can define its Powerball variant by simply adding a power coefficient to the gradient computation in L-BFGS. The L-BFGS Powerball method is presented in Algorithm 1:

---
**Algorithm 1** L-BFGS Powerball method
---
1: $g_k = \nabla f(x(k))$, $q = \sigma(g_k)$
2: **for** $i = k-1, k-2, \ldots, k-m$ **do**
3: $\quad \alpha_i = \rho_i s_i^T q$,
4: $\quad q = q - \alpha_i y_i$,
5: $\quad H_k^0 = y_{k-1}^T s_{k-1} / y_{k-1}^T y_{k-1}$,
6: $\quad z = H_k^0 q$.
7: **end for**
8: **for** $i = k-m, k-m+1, \ldots, k-1$ **do**
9: $\quad \beta_i = \rho_i y_i^T z$,
10: $\quad z = z + s_i(\alpha_i - \beta_i)$.
11: **end for**
12: Stop with $H_k g_k = z$
---

## 5. Experiments

To evaluate the Powerball methods, we collected three datasets, which are listed in Table 1. RCV1 is a Reuters news classification dataset[1]. KDD10 is sampled from the KDD Cup 2010[2], whose goal is to measure students' performance. CTR is a sampled ad click-through rate dataset[3].

We used the logistic regression with $\ell_2$-regularization as the objective function. Given a list of example pairs $\{y_i, x_i\}_{i=1}^{n}$, the goal is to solve the following minimization problem

$$\min_{w} \sum_{i=1}^{n} \log(1 + \exp(-y_i \langle x_i, w \rangle)) + \lambda \|w\|_2^2. \quad (5.1)$$

---
[1] http://about.reuters.com/researchandstandards/corpus/
[2] https://pslcdatashop.web.cmu.edu/KDDCup/
[3] http://data.dmlc.ml

| name  | # examples         | # features         | # nonzero entries  |
|-------|--------------------|--------------------|--------------------|
| RCV1  | $2.0 \times 10^4$  | $4.7 \times 10^4$  | $1.5 \times 10^6$  |
| KDD10 | $2.0 \times 10^5$  | $6.4 \times 10^5$  | $7.4 \times 10^6$  |
| CTR   | $2.2 \times 10^5$  | $6.2 \times 10^5$  | $1.3 \times 10^7$  |

Table 1: Standard benchmark datasets for classification.

We used $\lambda = 1$ for KDD10 and CTR while $\lambda = 0$ for RCV1.

Both the gradient descent and L-BFGS (Liu and Nocedal, 1989) are compared with the gradient Powerball method and L-BFGS Powerball method from the same initial conditions which are randomly chosen. The step size in both methods is chosen by standard backtracking line search. The weight $w$ is initialized according to a normal distribution $\mathbf{N}(0, 0.01)$. We repeat each experiment 10 times and report the averaged results. The codes are available from http://yy311.github.io/software.html for the readers to reproduce the experimental results.

We first study the effect of varying $\gamma$. We choose four $\gamma$ values from a set $\{1, 0.7, 0.4, 0.1\}$, where for $\gamma = 1$ we obtain standard gradient descent. The convergence of different optimization algorithms for each $\gamma$ are shown in Fig. 1. As can be seen, when a $\gamma < 1$ is applied to the gradient in every steps, it can significantly accelerate the convergence as compared with the standard gradient descent method. Especially, on both KDD10 and CTR datasets, less than 10 iterations with $\gamma = 0.1$ can result an objective value even less than the one for gradient descent method with 100 iterations. The results for L-BFGS Powerball method comparison ($m = 5$) are shown in Fig. 2, which are similar to the observations for gradient Powerball method.

## 6. Discussion

It is generally known that dynamical systems (Sastry, 1999) can offer new insight to optimization methods (Lessard et al., 2016; Su at al., 2015; Krichene et al., 2015) by viewing optimization algorithms as evolution of dynamical systems. Using intuition from finite-time stability of ordinary differential equations (Bhat and Bernstein, 1997), we generalize the idea to the discrete schemes for optimization and demonstrate that empirically the proposed methods can accelerate the process in the initial iterations. When it comes to large-scale optimization problems, initial iterations are crucial given computation constraints.

## 7. Acknowledgments

We would like to thank Dr. Vassilis Vassiliadis, Dr. Benjamin Recht, and Dr. Stephen Boyd for insightful discussions.

## Appendix A. Proof of Proposition 1

**Lemma 1.** *((Bhat and Bernstein, 1997), Theorem 1) Suppose that a function $V(t) : [0, \infty) \to [0, \infty)$ is differentiable (the derivative of $V(t)$ at 0 is defined as its Dini upper derivative), such that $\frac{dV(t)}{dt} + KV^\gamma(t)$ is negative for all $t$, for some constant $K > 0$ and $0 < \gamma < 1$. Then $V(t)$ will reach zero at finite time $t^* \leq \frac{V^{1-\gamma}(0)}{K(1-\gamma)}$, and $V(t) = 0$ for all $t \geq t^*$.*

*Proof.* Let $f(t)$ satisfy the following ODE $\frac{df(t)}{dt} = -Kf^\gamma(t)$. Given any initial value $f(0) = V(0) > 0$, its unique solution is

$$f(t) = \begin{cases} \left(-K(1-\gamma)t + V^{1-\gamma}(0)\right)^{\frac{1}{1-\gamma}}, & t < \frac{V^{1-\gamma}(0)}{K(1-\gamma)} \\ 0, & t \geq \frac{V^{1-\gamma}(0)}{K(1-\gamma)} \end{cases}.$$

Since $V(0) = f(0)$, by the Comparison Principle of differential equations in (Bhat and Bernstein, 1997), we have $V(t) \leq f(t)$, $t \geq 0$. Hence, $V(t)$ will reach zero in time $\frac{V^{1-\gamma}(0)}{K(1-\gamma)}$. Since $V(t) \geq 0$ and $\frac{dV(t)}{dt} \leq 0$, $V(t)$ remains 0 once convergence. □

Next, we shall construct a Lyapunov function for eq. (2.1), which has a similar property in Lemma 1. Let $y_i = \frac{\partial f(x)}{\partial x_i}$, and consider a nonnegative function $V(t) = \frac{1}{\gamma+1} \sum_{i=1}^n |y_i|^{\gamma+1}$. If we take the derivative of $V(t)$ with respect to $t$, then we have

$$\frac{\partial V(t)}{\partial t} = \sum_{i=1}^n \frac{\partial V(t)}{\partial y_i} \frac{\partial y_i}{\partial t} \stackrel{(a)}{=} \sum_{i=1}^n \text{sign}(y_i)|y_i|^\gamma \left(\sum_{j=1}^n \frac{\partial y_i}{\partial x_j} \frac{\partial x_j}{\partial t}\right)$$

$$= -\begin{bmatrix} \text{sign}(y_1)|y_1|^\gamma & \ldots & \text{sign}(y_n)|y_n|^\gamma \end{bmatrix} H(y_i)$$

$$\begin{bmatrix} \text{sign}(y_1)|y_1|^\gamma & \ldots & \text{sign}(y_n)|y_n|^\gamma \end{bmatrix}^T$$

$$\stackrel{(b)}{\leq} -m \sum_{i=1}^n |y_i|^{2\gamma} \stackrel{(c)}{\leq} -m(\gamma+1)^{\frac{2\gamma}{\gamma+1}} V^{\frac{2\gamma}{1+\gamma}}(t). \tag{A.1}$$

Equality (a) is due to the fact that $\forall i$, $\frac{\partial |y_i|^{\gamma+1}}{\partial y_i} = (\gamma+1)\text{sign}(y_i)|y_i|^\gamma$. Inequality (b) is due to the Hessian $H \triangleq [\frac{\partial^2 f}{\partial x_i \partial x_j}] \succeq mI$ for any strongly convex function $f$. Inequality (c) holds using the fact that $\sum_{i=1}^n |y_i|^{2\gamma} \geq (\sum_{i=1}^n |y_i|^{\gamma+1})^{\frac{2\gamma}{\gamma+1}}$, $\forall \gamma \in (0, 1)$.

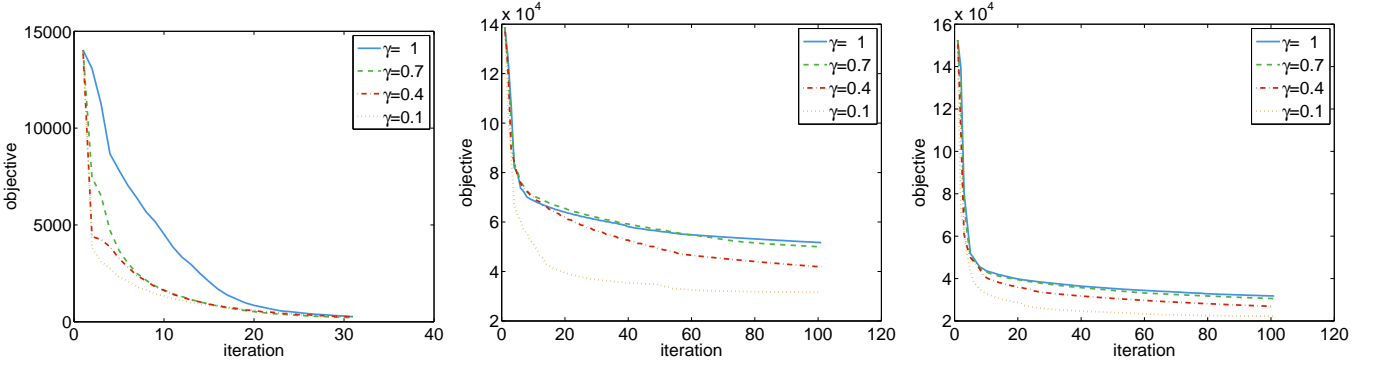

Figure 1: We apply Gradient Powerball method ($\gamma < 1$) and gradient descent method ($\gamma = 1$) to minimize eq. (5.1) on three datasets. Left: RCV1, middle KDD10, right: CTR. We observe that Gradient Powerball method with $\gamma$ less than 1 can significantly accelerate the convergence. Especially, on both KDD10 and CTR datasets, the objective value of eq. (5.1) that Gradient Powerball method achieved using 10 iterations (with $\gamma = 0.1$) would require 100 iterations for the standard gradient descent method.

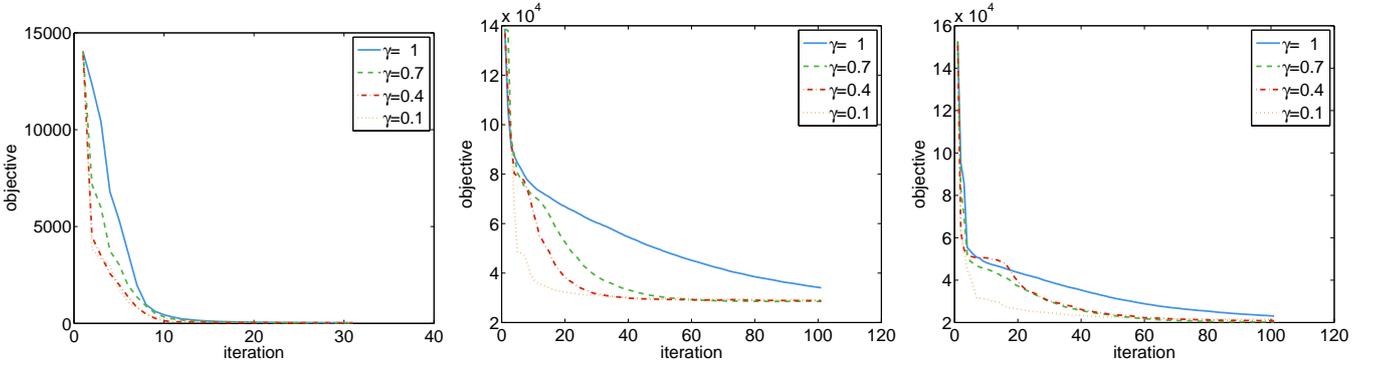

Figure 2: We apply L-BFGS Powerball method ($\gamma < 1$) and L-BFGS ($\gamma = 1$) to minimize eq. (5.1) on three datasets. We observe a similar result as the comparison of the gradient Powerball method with the gradient descent method. Left: RCV1, middle News20, right: CTR.

Using Lemma 1, eq. (A.1) implies that there exists $T = \frac{V^{\frac{1-\gamma}{1+\gamma}}(0)}{m} \frac{(\gamma+1)^{\frac{1-\gamma}{1+\gamma}}}{1-\gamma}$, $\forall \gamma \in (0, 1)$ such that $V(t) = 0$ when $t \geq T$. This implies that the system's state is at its equilibrium.

**Appendix B. Proof of Proposition 2**

Consider a nonnegative function $V(t) = \frac{1}{\gamma+1} \sum_{i=1}^{n} |\nabla f_i(x)|^{\gamma+1}$, similar to the proof of Proposition 1, if we take the derivative of $V(t)$ with respect to $t$ and then we have that, for all $t \geq 0$ and $\gamma \in (0, 1)$,

$$\frac{\partial V(t)}{\partial t} = -\left\|\begin{bmatrix}|\nabla f_1(x)|^\gamma & \ldots & |\nabla f_n(x)|^\gamma\end{bmatrix}\right\|_2^2 \leq -(\gamma+1)^{\frac{2\gamma}{\gamma+1}} V^{\frac{2\gamma}{1+\gamma}}(t).$$

Applying Lemma 1 leads to the result.

**Appendix C. Proof of Theorem 1**

*Proof.* Denote by $x^\star$ the minimizer and $f^\star = f(x^\star)$. For brevity, write $x_k = x(k)$ for all $k \geq 0$. Then, by the $L$-Lipschitz continuity of $\nabla f$ (Nesterov, 2014) and (3.1),

$$f(x_{k+1}) \leq f(x_k) + \nabla f(x_k) \cdot (x_{k+1} - x_k) + \frac{L}{2}|x_{k+1} - x_k|^2$$
$$= f(x_k) - \alpha_k \nabla f(x_k) \cdot \sigma(\nabla f(x_k)) + \frac{L}{2}\alpha_k^2 |\sigma(\nabla f(x_k))|^2.$$

Let $\alpha_k = \frac{\nabla f(x_k) \cdot \sigma(\nabla f(x_k))}{L|\sigma(\nabla f(x_k))|^2} > 0$ (this holds if $x_k \neq x^\star$). Then

$$f(x_{k+1}) \leq f(x_k) - \frac{(\nabla f(x_k) \cdot \sigma(\nabla f(x_k)))^2}{2L|\sigma(\nabla f(x_k))|^2}. \quad (C.1)$$

Note that

$$\frac{(\nabla f(x_k) \cdot \sigma(\nabla f(x_k)))^2}{|\sigma(\nabla f(x_k))|^2} = \frac{(\sum_{i=1}^{n} |(\nabla f)_i(x_k)|^{\gamma+1})^2}{\sum_{i=1}^{n} |(\nabla f)_i(x_k)|^{2\gamma}}$$
$$\geq \frac{\sum_{i=1}^{n} |(\nabla f)_i(x_k)|^2}{n} = \frac{|\nabla f(x_k)|^2}{n}. \quad (C.2)$$

The previous inequality follows from

$$n(\sum_{i=1}^{n} |y_i|^{\gamma+1})^2 \geq n \sum_{i=1}^{n} |y_i|^{2\gamma}|y_i|^2 \geq (\sum_{i=1}^{n} |y_i|^{2\gamma})(\sum_{i=1}^{n} |y_i|^2),$$

for $0 \leq \gamma \leq 1$ and a vector $y \in \mathbb{R}^n$, where the last inequality is Chebyshev's sum inequality. By strong convexity of $f$ (Nesterov, 2014),

$$|\nabla f(x)|^2 \geq 2m(f(x) - f^\star), \quad \forall x \in \mathbb{R}^n. \quad (C.3)$$

Combining (C.1), (C.2), and (C.3) gives

$$f(x_{k+1}) - f^\star \leq [1 - \frac{1}{n} \cdot \frac{m}{L}](f(x_k) - f^\star). \quad (C.4)$$

This shows that the Powerball scheme converges linearly at a rate $(1 - O(\frac{m}{L}))^k$, by the choice of time-varying steps above. □